# Low-Symmetry Nanophotonics

Alex Krasnok[1], Andrea Alú[1,2]

[1]*Photonics Initiative, Advanced Science Research Center, City University of New York, New York, NY 10031, USA*

[2]*Physics Program, Graduate Center, City University of New York, New York, NY 10016, USA*

## Abstract

*Photonics and optoelectronics are at the foundations of widespread technologies, from high-speed Internet to systems for artificial intelligence, automotive LiDAR, and optical quantum computing. Light enables ultrafast speeds and low energy for all-optical information processing and transport, especially when confined at the nanoscale level, at which the interactions of light with matter unveil new phenomena, and the role of local symmetries becomes crucial. In this Perspective, we discuss how symmetry violations provide unique opportunities for nanophotonics, tailoring wave interactions in nanostructures for a wide range of functionalities. We discuss geometrical broken symmetries for localized surface polaritons, the physics of moiré photonics, in-plane inversion symmetry breaking for valleytronics and nonradiative state control, time-reversal symmetry breaking for optical nonreciprocity, and parity-time symmetry breaking. Overall, our Perspective aims at presenting under a unified umbrella the role of symmetry breaking in controlling nanoscale light, and its widespread applications for optical technology.*

The invention of fiber optics, in which light is trapped in the lateral direction through total internal reflection, has enabled high-speed and efficient long-distance optical communications. Highly efficient light propagation and the availability of advanced repeaters, amplifiers and remarkable bandwidth have made this technology ideal for broadband connections well above 1 Gbps. Fig. 1a shows a map of intercontinental submarine fiber-optic connections, through which the vast majority of information is transmitted today. It is difficult to overestimate the value of this technology, which emerged in the middle of the last century but found its use in vital applications much later. However, this technology faces a significant bottleneck: when information is received in the form of optical pulses at one of the data centers, it must be converted into an electrical form to be processed electronically. Electronic components run much slower than light and are characterized by large energy consumption. According to the U.S. Department of Energy, some of the world's largest data centers contain tens of thousands of electronic devices and use over 100 megawatts (MW) of power - enough energy to power about 80,000 U.S. households[1,2]. Most of this energy goes into heat and it is eventually lost. All-optical signal processing and transport holds the promise for large improvements in terms of speed and efficiency of today's data centers. Indeed, the idea of all-optical computing has been gaining traction in recent years[3]. Unlike electrons, photons are bosons and, as such, they do not interact with each other in linear media, which offers challenges in the context of data processing and computing, but also various new opportunities. Indeed, the bosonic nature of photons allows them to maintain a quantum state for a long time, much longer than the typical quantum coherence time in solid-state quantum systems, including quantum dots (QDs), defect centers (for example, NV centers) and quantum circuits with Josephson junctions. The robustness of the quantum state of light has enabled applications in quantum cryptography[4–7] and emerging optical quantum computing[8–10]. In addition, the large size of photons enables quantum phenomena at microscopic and even macroscopic scales, which greatly simplifies their use and scalability.

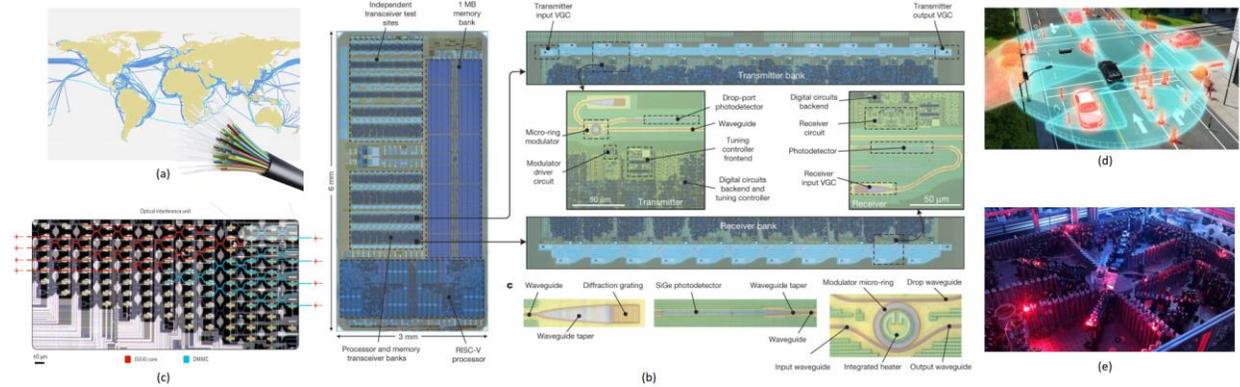

**Figure 1. Advanced and nascent applications of photonics**. (a) Map of intercontinental submarine fiber-optic connections. (b) Schematics of a hybrid electro-optical computers[11]. (c) Neuromorphic photonics for deep learning applications[12]. (d) LiDAR technology. (e) Quantum optical computers[9].

Using hybrid systems that combine electronic and photonic components in tandem appears to be a natural approach to leverage the advantages of both systems and it is likely to be the next technological breakthrough towards all-optical computers. Fig. 1b shows an example of one such hybrid electronic photonic system integrated in a single chip, incorporating over 70 million transistors and 850 photonic components that work together to provide logic, memory and interconnect functions[11]. Our PCs are likely to contain such hybrid electro-optical chips to improve performance in highly demanding computing and rendering functionalities.

Photonics intends to revolutionize vital areas other than information technology. One example is non-digital computing models, such as neural networks and neuromorphic computing[12–15], where ultra-fast hardware based on optical integrated circuits has made it possible to create a new class of information processing machines with applications derived from training[16] and nonlinear programming[17] to intelligent signal processing[18] and medical diagnosis. Fig. 1c exemplifies such a system[12], consisting of a programmable nanophotonic processor containing a cascaded array of 56 programmable Mach-Zehnder interferometers (MZI meshes) in a silicon photonic integrated circuit. In such integrated optical processors, connections between pairs of artificial neurons are described by a scalar synaptic weight, so that the layout of interconnections can be represented as a matrix-vector operation, where the input to each neuron is the dot product of the output from connected neurons attenuated by a weight vector. Optical signals can be multiplied by transmission through tunable waveguide elements. These neural networks require high scalability and, as such, relatively long-range connections to perform nontrivial distributed processing, making the use of integrated photonics unavoidable. These optical platforms will be useful not only for information processing but also for other relevant applications, including robot control, mathematical programming, and testing neurobiological hypotheses[19,20].

Optical radars or LiDARs (Light Detection and Ranging) are another great examples of how light technology is changing our lives and technologies, enabling or contributing to an infinite range of technologies: laser guidance, surveying, archeology, geology, seismology, atmospheric physics[21,22]. This technology uses laser light to measure range, such as distance to resting or moving objects (Fig. 1d), attitude to the Earth etc. Finally, programmable photonic circuits are considered to be promising for room temperature quantum computers, the area that recently got a significant boost after reporting several groundbreaking results[9,23]. Fig. 1e shows such an optical quantum computer, which was reported last year as being capable of sampling bosons by sending several (40–70 in this example) indistinguishable photons through the ports of a linear interferometer consisting of multiple mirrors and beam splitters. When photons pass through the system, their position and trajectory change randomly. About ten years ago, it was argued that a classical computer would not be able to compute the output of such a system[24]. Thus, boson sampling could be used to demonstrate the advantage of a quantum computer over classical computing systems. The progress in optical quantum computing operating at room temperature has been very exciting to observe.

While light potentially offers terrific opportunities for these applications in terms of speed and energy savings, the main challenge is that these benefits can be unveiled only under the assumption that optical devices can be drastically miniaturized. The size to which a free photon can be squeezed into is bounded by the diffraction limit, $\propto \lambda_0 / n$, where $\lambda_0 = 2\pi c/\omega$ is the free space wavelength, $\omega = 2\pi f$ stands for angular frequency, $c$ is the speed of light, and $n$ is the refractive index of the material in which the photon is localized. For the telecom wavelength $\lambda_0 = 1{,}550$ nm and refractive index of silica $n = 1.52$, this yields ~1 µm for the photon size, which is roughly three orders of magnitude greater than the electron size (de Broglie wavelength). Common photonic elements such as lenses and optical fibers require a size on the order of wavelength (a few microns) due to the diffraction limit. Technically speaking, the (normalized) mode volume ($V$) of such photonic modes in any dielectrics is bounded $V/V_\lambda \geq 1$, where $V_\lambda = (\lambda/n)^3$. Even cavities with large Q-factors or those made of high refractive index (for example, n ≃ 4 for Si in visible and IR range) do not allow going below this diffraction limit, Fig. 2a. Also, an attempt to compress the mode in the resonator leads to the fact that most of the field is localized in its material, resulting in a reduction of the Q-factor due to increased dissipation. As a result, a tradeoff between the Q-factor and the mode volume appears, Fig. 2a. The diffraction limit leads to the fact that photonic systems require $10^4$ more area for an equivalent device functionality[3,25]. In addition, the speed and energy consumption of a device are fundamentally limited by its size, implying that only drastic miniaturization well below the diffraction limit, and towards quantum scales, can realize systems that can compete and even outperform electronic systems.

Compressing light to the nanometer scale is possible by strongly coupling light with matter, as it happens in polaritonic systems relying on, e.g., plasmons (plasmon polaritons), phonons (phonon polaritons), magnons (magnon polaritons), etc. Polariton modes can have a wavelength much shorter than light, which leads to their tight localization. For example, oscillations of electrons at a metal-dielectric interface support surface plasmon polaritons (SPPs) with a large momentum mismatch and localization limited only by material losses and surface imperfections. Plasmon-polariton modes in nanoparticles or metal waveguides also find several important applications in sensing, light-harvesting, nonclassical light sources, and communications[26–34]. Fig. 2b shows a state-of-the-art device comprising an SPP-based interferometer (yellow) driven by an electro-optical element (modulator) controlled by electronics. Due to the tiny size of this device and the use of highly localized SPPs, the modulator exhibits a data modulation rate of 120 Gbps with a monolithically integrated transmitter, which is far superior to any electronic counterpart. There are other varieties of polaritons, including those formed by excitons in semiconductors, atomic vibrations in polar insulators, Cooper pairs in superconductors, and spin resonances in (anti) ferromagnets. They cover a wide range of the electromagnetic spectrum, from microwave to ultraviolet[35].

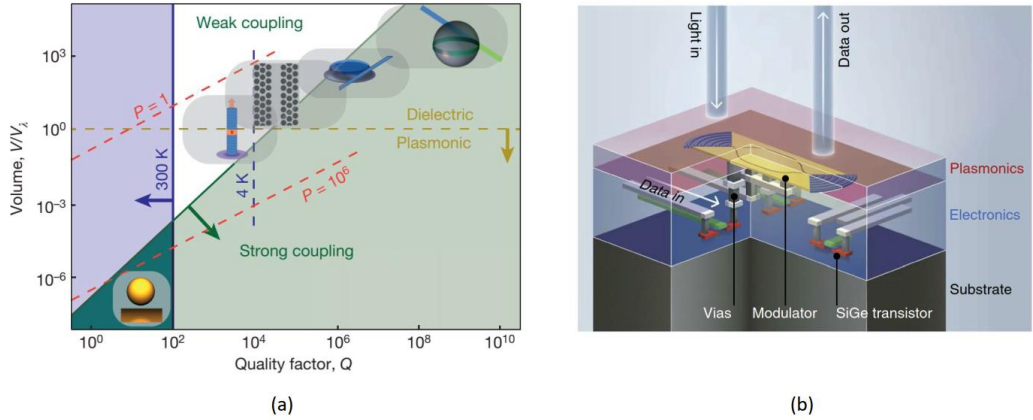

**Figure 2**. (a) Trade-off between quality factor of a nanocavity and its effective volume, $V/V_\lambda$ showing strong-coupling (green arrow), room-temperature (blue arrow), and plasmonic (orange arrow) regimes. The icons show realizations of each type of nanocavity: from right, whispering gallery spheres, microdisks, photonic crystals, micropillars, and nanoparticle-on-mirror geometry. Purcell factors (P) show emission-rate enhancements.[36] (b) Ultra-compact plasmonic Mach-Zehnder modulator with a footprint of $29 \times 6$ μm$^2$ for co-integration with electronics[37], demonstrating a 120 GBps rate of data modulation with a monolithically integrated transmitter.

As we discuss in this Perspective, when light is squeezed down to such small dimensions, the local symmetries of nanophotonic structures play a fundamental role. Playing with anisotropy enables interesting regimes of operation and novel functionalities, including hyperbolic modes and topological transitions, the novel physics of moiré patterns, spin-orbit locking, enhanced valley polarization and chirality. The rotation and translation symmetries of optical structures enable nonradiative and nonscattering modes with unbounded Q-factors. Time-reversal symmetry breaking enables nonreciprocity and isolating devices. Herein we discuss how nanophotonics with broken symmetries heralds a new era of light technologies.

**A. Hyperbolic materials and metasurfaces**

The first type of broken symmetry we discuss in this paper is the in-plane anisotropy of flat materials and metasurfaces. Metasurfaces, the two-dimensional analogue of metamaterials, are artificial structures composed of subwavelength, usually resonant elements (metaatoms) tailored and arranged to achieve a certain functionality like control over the light propagation, reflection, and refraction[38–52]. Unlike metamaterials, metasurfaces are flat, fully compatible with modern manufacturing technology, easy to integrate into on a chip while retaining most of the functions of three-dimensional metamaterials. Optical metasurfaces offer an alternative approach for the implementation of optical components. Recent advances have increased their efficiency and functionality, allowing metasurface-based diffractive optical components to have performance comparable or superior to conventional optical components[53–55]. The main advantage of metasurfaces stems from the ability to create complex planar optical systems consisting of lithographically stacked electronic and metasurface layers. The resulting optical system is lithographically aligned, eliminating the need for post-fabric alignment. According to the World Economic Forum, planar

MS optics is about to enter the top ten emerging technologies and, according to forecasts by Lux Research, will bring $10 billion to the market by 2030[56].

Metasurfaces allow engineering tightly confined surface polaritonic waves enabling extremely high light-matter interactions. Unlike bulk waves, surface polaritons (SPs) in metasurfaces have evanescent fields in the direction perpendicular to the surface ($\text{Im}(k_z) > 0$), which provides greater confinement, enhanced local density of states and more robustness. A mathematical model of these structures is based on their complex surface conductivity tensor, $\overline{\overline{\sigma}} = (\{\sigma_{xx}, \sigma_{xy}\}, \{\sigma_{yx}, \sigma_{yy}\})$, which can be written in diagonal form in the main coordinate system as (unless magneto-optical effects or more complex phenomena arise)

$$\overline{\overline{\sigma}} = \begin{pmatrix} \sigma'_{xx} & 0 \\ 0 & \sigma'_{yy} \end{pmatrix}. \qquad (1)$$

In the overwhelming majority of natural thin materials (including graphene) and artificial metasurfaces, the diagonal components have the same sign, $\text{sgn}(\sigma'_{xx}) = \text{sgn}(\sigma'_{yy})$, while the absolute value can be different, resulting in an elliptical topology of the corresponding band diagram representing the modal resonances supported by the surface. For instance, SPPs in pristine graphene follow a band diagram with (isotropic) circular topology[57].

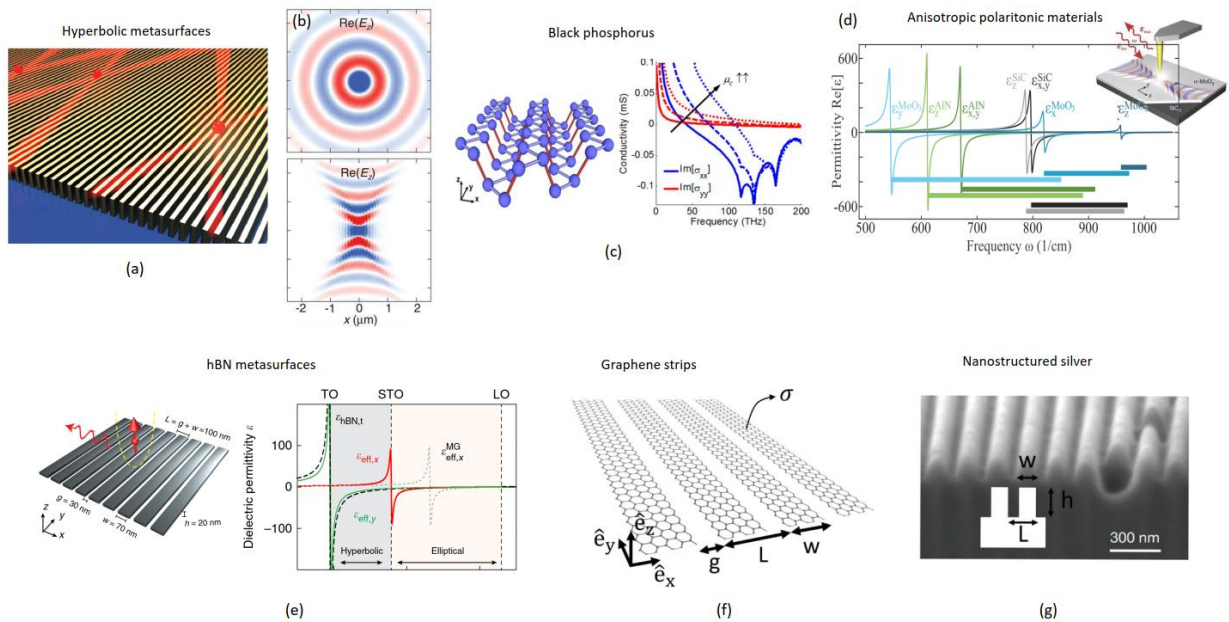

**Figure 3. Hyperbolic natural and artificial materials.** (a) Hyperbolic metasurface. (b) Elliptical (top) and hyperbolic (bottom) profile of the surface-polaritonic waves. (c) Black phosphorous – natural hyperbolic material[64]. (d) Other examples of natural hyperbolic materials, anisotropic materials with phonon-polaritonic resonance[65]. (e) Hyperbolic material can be made by introducing anisotropy, making a grating on a surface of hBN [63]. (f) (g) Hyperbolic metasurfaces based on patterning graphene[61] and metallic (Ag) slab[62].

Of particular interest are extremely anisotropic responses that arise in hyperbolic metasurfaces (HMTSs) characterized by different signs of the conductivity components, $\text{sgn}(\sigma'_{xx}) \neq \text{sgn}(\sigma'_{yy})$ [57]. Probably, the simplest physical realization of such a surface is a periodic array of metal wires or rods with subwavelength granularity, Fig. 3a. In this case, the dispersive conductivity along the wires is metallic, while across it is dielectric. The conductivity components are dispersive, and the structure can demonstrate a so-called topological transition, going from an elliptic to a hyperbolic regime with an abrupt change in the angular dispersion of surface polaritons (SPs) and the local density of states (LDOS), Fig. 3b. The hyperbolic operation regime has been demonstrated in several material platforms, from black phosphorus[58] (Fig. 3c) and other 2D materials[59] (Fig. 3d), nanostructured van der Waals materials[60] (Fig. 3e), graphene nanoribbons[61] (Fig. 3f) and silver gratings[62] (Fig. 3g). For example, deeply subwavelength graphene or hBN nanoribbon arrays can form HMTSs, featuring extreme surface anisotropy [60,61,63].

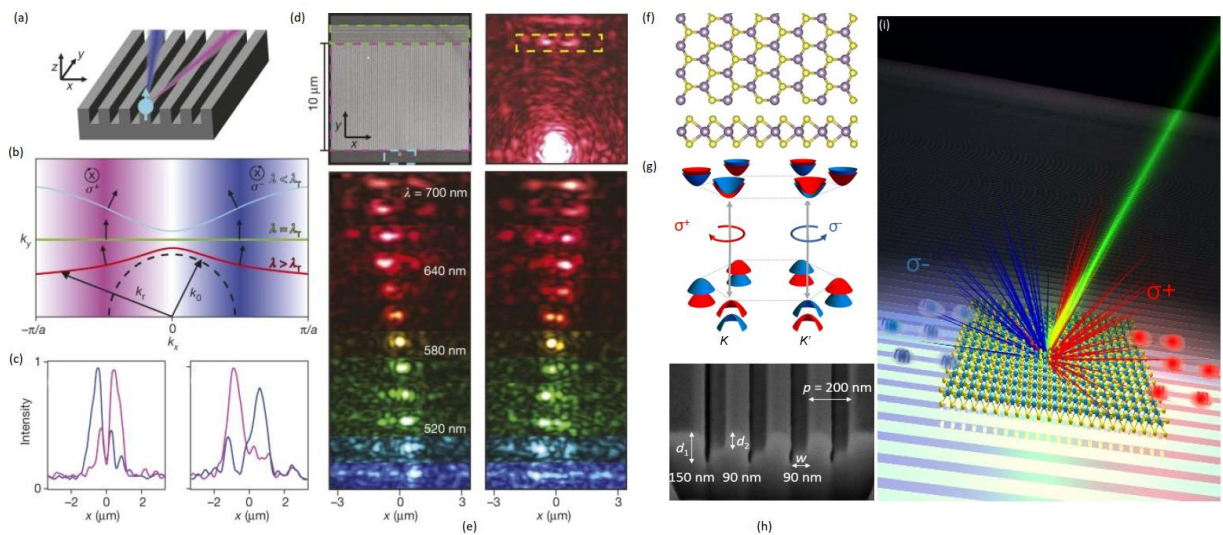

**Figure 4**. (a) Schematic of the plasmonic spin-Hall effect. In a silver/air grating, SPPs with different helicities propagate into distinct spatial directions[62]. (b) Direction of the group velocity (black arrows) is perpendicular to the isofrequency contours at a given dispersion regime: elliptic (red), diffractionless (green), and hyperbolic (cyan). (c) Light intensities measured at the out-coupling structures for $\sigma^+$- (magenta) and $\sigma^-$- (blue) polarized light at $\lambda = 530$ nm (left) and $\lambda = 640$ nm (right). (d) SEM image of the device supporting plasmonic spin-Hall effect with an in-coupling structure (cyan rectangle), a silver/air grating (pink rectangle), and out-coupling cylinders (green rectangle). (right) Image under unpolarized laser excitation of the in-coupling structure. (e) Image from the out-coupling structures as a function of wavelength collecting only $\sigma^+$ (left) and $\sigma^-$ (right) polarized light. (f) Crystal structure of monolayer transition metal dichalcogenides. (g) Scheme of the band structure and selection rules at the boundary of the Brillouin zone. (h) SEM of asymmetric grooves. (i) Illustration of valley excitons separation in real space and polarized emission separation in momentum space[74].

HMTSs are promising for optical on-chip interconnects, sensing and quantum technologies. In recent work[62], a HMTS (Fig. 4a) exhibiting strong, dispersion-dependent spin-orbit coupling, enabling polarization- and wavelength-dependent routing of plasmonic SP and two-dimensional chiral optical components has been reported. The realized optical HMTS was made of single-crystal Ag by lithographic and etching techniques (Fig. 4d) and demonstrates negative refraction and diffraction-free propagation inherent to bulk metamaterials but on a 2D layout and with better energy consumption. The HMTS exhibits a topological transition from elliptical to hyperbolic mode as the operation wavelength varies (Fig. 4b). Based on this platform, various exotic phenomena can emerge, including the plasmonic spin-Hall effect[66–69], an analog of the electronic Rashba effect[70,71], and photonic spin-Hall effects[72,73]. In solid-state physics, spin-orbit coupling links the spin of a charge carrier with the direction of its propagation due to the lack of inversion symmetry of the crystal. Likewise, in optics the rotation of an electric field or polarization blocks the direction of propagation of light, causing the photonic spin-Hall effect. The plasmonic spin-Hall effect consists in locking of the propagation direction of surface plasmon-polaritons producing a hot spot on the opposite side of the structure at different locations depending on the polarization, Fig. 4c. When the operating wavelength changes, the HMTS changes its operating mode, which leads to the launch of SP waves of a different wavelength in a different direction, Fig. 4e.

In a related scenario, such plasmonic MS has been designed to sort in space opposite-handed $\sigma^+$ and $\sigma^-$ exciton polaritons arising in 2D transition metal dichalcogenides (2D TMDCs), a promising functionality for valleytronics and optoelectronics. 2D TMDCs possess a direct bandgap in the visible frequency region due to the significantly reduced dielectric screening effect. Similar to graphene, 2D TMDCs have a hexagonal crystalline lattice (Fig. 4f), resulting in six valleys of the hexagonal Brillouin zone, but because of the lack of inversion symmetry, their valleys are characterized by two sorts of opposite spins (+K and -K valleys, Fig. 4f). This feature offers a unique platform for controlling spin and valley degrees of freedom in valleytronics, where valley polarization can be used as an alternative low-energy information carrier[75]. However, the short depolarization time of the valleys and the fast lifetime of excitons in 2D TMDCs at room temperature prevent the use of pseudospins of valleys in practical on-chip applications. Resonant optical nanostructures have recently been proposed to enhance and separate excitons from opposite valleys to different directions[74]. In particular, it was demonstrated that valley-polarized excitons can be sorted and spatially separated at room temperature by coupling 2D $MoS_2$ to a designed asymmetric groove metasurface (Fig. 4h)[74]. The absence of in-plane mirror symmetry makes it possible to spatially separate excitons with opposite polarization that live in the plane of the 2D material (Fig. 4i). The approach can be extended to a wide range of material platforms, which makes it possible in practice to use the valley's degree of freedom for processing and storing information at room temperature.

The coupling of exciton-polaritons in other 2D materials with different structures supporting the optical spin Hall effect should be extensively investigated to improve photonic and optoelectronic devices and provide unprecedented opportunities arising from this hybrid combination of quantum, nanoscale and photonic engineering. Since this area is still in its infancy, we envision further leveraging strong spin-orbit interactions in 2D TMDCs, which couples the spin of carriers to the valley band index at the corners of the Brillouin zone, enabling manipulation of carrier spins via optoelectronic processes.

**B. Twistronics and moiré photonics**

It has recently been shown that the close coupling of two thin materials or metasurfaces supporting extreme anisotropymay provide a convenient way to drastically tune their optical properties and demonstrate emergent new optical phenomena. It is noteworthy that the physics of such systems can be related to the physics of moiré patterns. Moiré patterns occur when two or more periodic patterns are superimposed with a difference in lattice constants or relative spatial displacement, Fig. 5a. Due to their high sensitivity to mechanical distortion, rotations, and displacements, moiré effects have found various applications, for example, in strain analysis[76,77] and optical alignment[78–81]. More discussions of moiré phenomena can be found in previous reviews.[82,83]. Interest in moiré structures recently revived after it was discovered that superimposed and twisted two-dimensional van der Waals heterostructures (Fig. 5b) offer a new platform for advanced nanoelectronics in the emerging field of twistronics - manipulating the electron wavefunction and its transport via the interlayer twist. This elegant concept has spawned many applications, including unconventional superconductivity in twisted graphene double layers (tBLs), moiré excitons in semiconductor tBLs, and interlayer magnetism in 2D magnetic tBLs. 2D van der Waals heterostructures support remarkable half-matter-half-photon polaritons that allow light to be manipulated at the nano and even angstrom scales for future applications in quantum and nonlinear optics[84]. The dispersion of polaritons in these tBLs can be interestingly engineered via a twist.

Optical analogues of tBL, consisting of two superimposed metasurfaces, have also been proposed and demonstrated as an effective way to tune the interactions of light and matter. This approach deals only with macroscopic systems and does not rely on atomic lattice effects. Being based on the interaction of electromagnetic waves, it offers simpler analysis and optimization while still providing highly exotic emerging optical phenomena based on twistronic concepts. For example, Wu and Zheng reported chiral plasmonic tBLs by stacking of two metasurfaces (thin Au films with periodic nanohole arrays of triangular lattices)[85], Fig. 5c. Remarkably, the positive and negative values of the interlayer rotation angle $\theta$ led to moiré metasurfaces with left-handed (LH) and right-handed (RH) configuration, Fig. 5d.

The surface anisotropy of hyperbolic metasurfaces such as arrays of deep subwavelength graphene and hBN nanoribbons can be tuned using the concept of moiré physics. In recent work, G. Hu et al. laid the foundation for moiré hyperbolic metasurfaces by stacking hyperbolic polaritonic tBL, which offers extreme manipulation of light at the nanoscale through rotation[45]. Extreme dispersion engineering of plasmon polaritons in tBL hyperbolic metasurfaces, each made of densely packed graphene nanoribbons (Fig. 5e) has been unveiled. Depending on the rotation angle, the supported hybridized polaritons undergo a transition from open hyperbolic to closed elliptical dispersion, Fig. 5f. At this specific angle, the system undergoes a photonic topological transition, analogous to the Lifshitz transition in electronics[86], with isofrequency contours simulating the Fermi surface for electrons[87]. The transition is characterized by the topological integer number of anti-crossing points, defined by crossing the two dispersion lines of the individual monolayers in momentum space. Near critical topological transition angles, the dispersion necessarily becomes flattened, supporting a highly collimated and diffractionless polariton transport, i.e., the polariton canalization. This angle was called "photonic magic angle,"[88–90] in analogy to solid-state magic angles where a flat Fermi surface is achieved in tBL graphene.

The idea was experimentally realized using 2D α-MoO$_3$ tBL[89], Fig. 5g. 2D α-MoO$_3$ is a van der Waals nanomaterial naturally endowed with in-plane hyperbolic phonon polaritons[91,92]. Using real-space nanoimaging, the hyperbolic to elliptical topological transition triggered by a change in the twist angle between the two layers was observed, and correspondingly canalized polaritons at the photon magic angles emerged over a wide range of frequencies. Fig. 5i shows the change in electric field pattern from hyperbolic to elliptical with an increase in the twist angle. The topological transition from open to closed dispersion occurs at the critical transition angle. At this twist angle, the dispersion flattens and the field becomes highly directional, reminiscent of the flat Fermi surface that is responsible for magic-angle superconductivity in bilayer graphene[93,94]. These results pave the way for extreme photonic dispersion engineering and polariton transport control based on a twisted stack of 2D materials with possible light manipulation applications at ultimately small dimensions and advanced quantum optics.

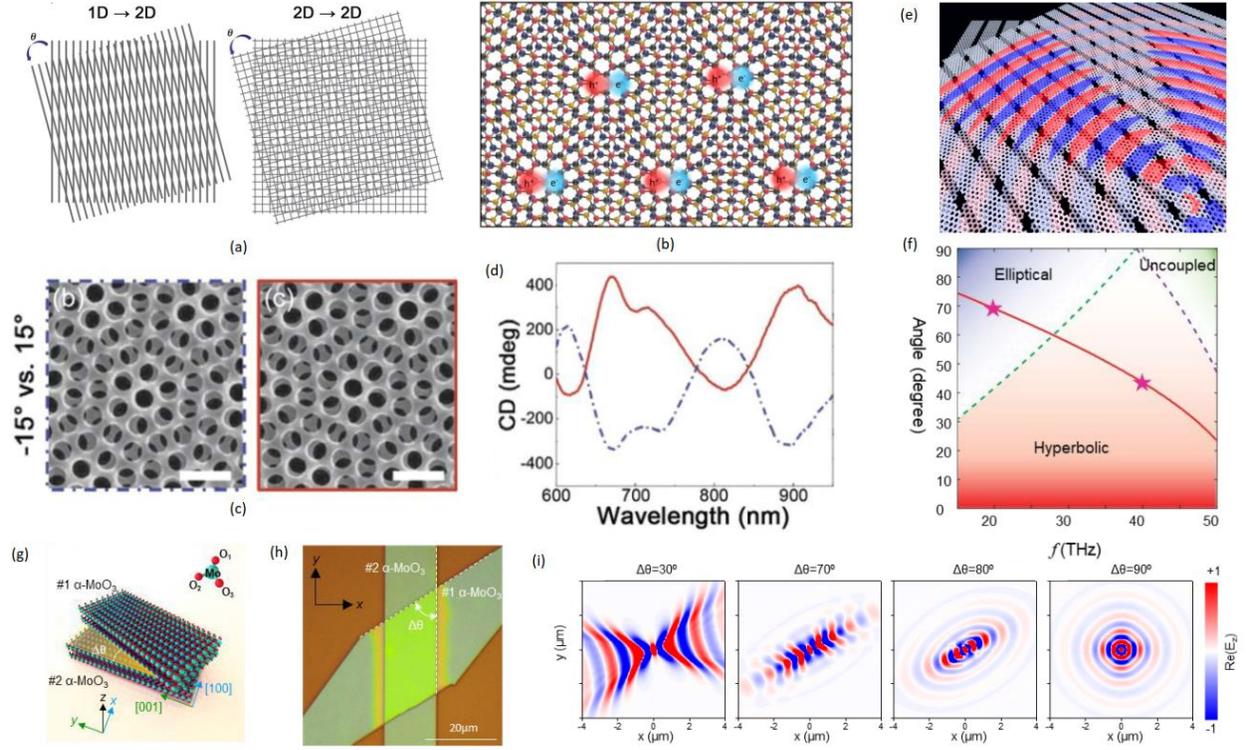

**Figure 5. Twistronics and moiré optics.** (a) Formation of 2D moiré pattern from two layers of 1D (left) or 2D (right) periodic structures at a relative in-plane rotation angle ($\theta$). (b) Moiré patterns in van der Waals heterostructures of dissimilar 2D materials. (c,d) Optical chirality in moiré chiral metamaterials: (c) SEM of moiré chiral metamaterials. Scale bar: 1 µm. Triangular-latticed constant: 500 nm; diameter of nanoholes: ≈350 nm. (d) Measured CD spectra of the left-handed (dashed) and right-handed (solid) moiré chiral metamaterials[85]. (e,f) Moiré hyperbolic metastructure composed of two coupled uniaxial graphene metasurfaces: (e) schematic view; (f) topological transition regions as a function of frequency and rotation angles[45]. (g-i) Topological polaritons in twisted α-MoO$_3$ bilayers: (g) schematic of tBL α-MoO3; (h) optical image of a tBL α-MoO3 structure; (i) electric field patterns of the tBL α-MoO3 structure for different relative twist angle[89].

## C. Chirality

Another type of structural asymmetry, chirality, implies that the resulting system does not obey *mirror symmetry*, Fig. 6a. Chirality is well known to lead to optical activity (rotation of the polarization of the electromagnetic field) in various optical settings. Chiral media demonstrate different absorption for left- and right-polarized light and this difference in absorption is called circular dichroism, $CD \propto A_{LCP} - A_{RCP}$. The larger CD, the stronger the optical activity effect. Chirality is a ubiquitous phenomenon in nature and many biomolecules without inversion symmetry, such as amino acids and sugars, are chiral molecules. Measuring and controlling molecular chirality with high precision down to the atomic scale is very important in physics, chemistry, biology and medicine. The first experiment with tailored materials for optical activity effect was conducted by J.C. Bose in 1898[95] in millimeter waves. Nowadays, circular polarization is obtained by stacking a linear polarizer with bulky quarter-wave plates made by birefringent materials, Fig. 6b. This configuration is based on a conventional polarimeter, where the transmission depends on the concentration of optically active materials. This configuration, however, cannot be easily integrated into nanophotonic systems and is inherently narrow bandwidth because conventional quarter-wave plates rely on wavelength-dependent birefringence phenomenon. This effect is also at the heart of our LCD screens and three-dimensional display technologies[96]. In pharmacology, this effect is used to detect and separate enantiomeric drugs when one chirality forms a powerful drug, while the other can cause serious side effects[97,98]. However, by themselves, chiral molecules usually have extremely small CDs, in the range of only a few tenths of a milli-degree, reaching a maximum in the violet and ultraviolet parts of the spectrum due to the size of the molecules. Consequently, conventional CD measurements require large integration times (a few tens of mins) to resolve these small signals, motivating researchers to develop systems capable of detecting small amounts of enantiomers.

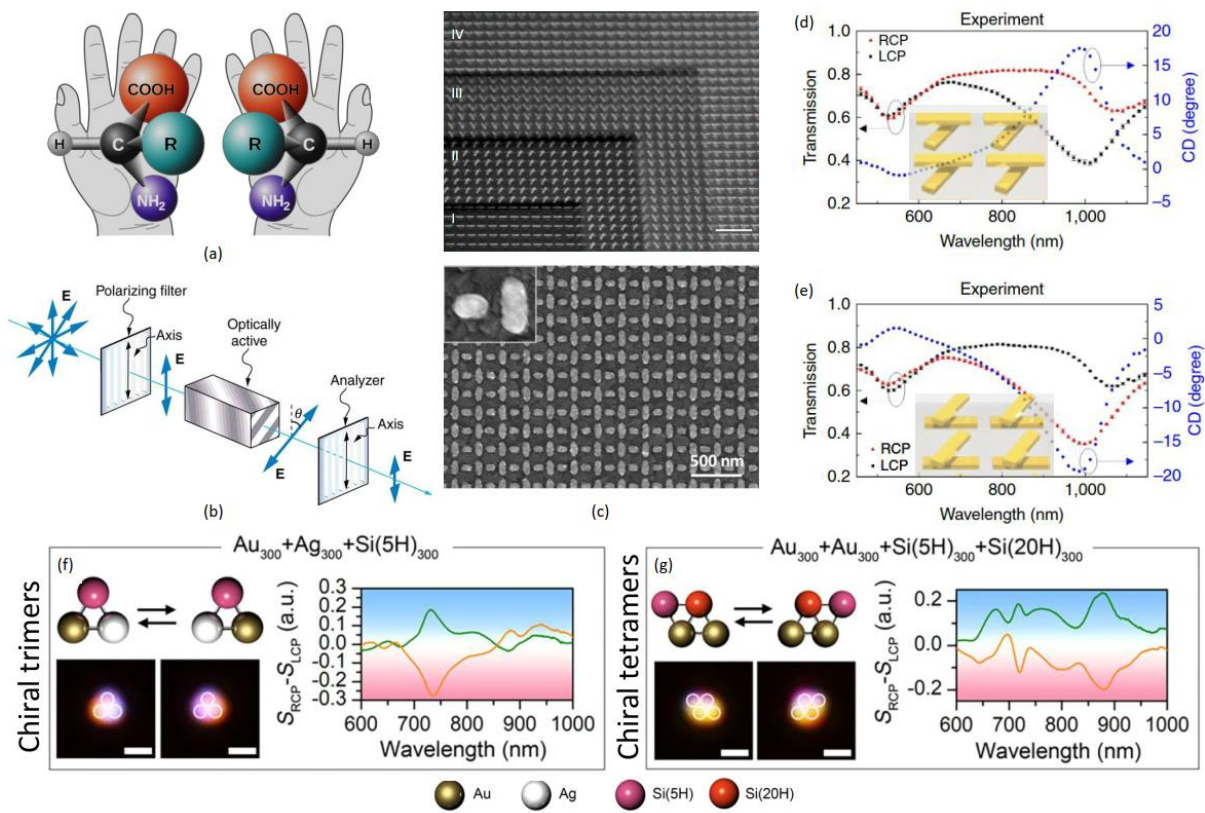

**Figure 6. Metastructures for enhanced chirality and optical activity.** (a) Enantiomers of a generic amino acid are chiral because they do not superimpose with their reflection in a mirror. (b) Polarimeter: the transmission depends on the concentration of optically active materials. Widely used in medicine, biology, pharmacology and industry. (c) Sophisticated 4-layer twisted metamaterial (top)[111] and metasurface (bottom)[110] for an enhanced chiral response. (d,e) Measured transmission (left y-axis) of the twisted metamaterial with RCP (red curve) and LCP (black curve) excitation, and its extracted circular dichroism (CD) in degrees (blue curve, right y-axis) for the +60º (d) and -60º (e) interlayer rotations. (f,g) Reconfigurable chiral meta-molecules [112]: (f,g) Schematic, optical images and differential scattering spectra of chiral trimers composed of (s) an AuNP, an 6 AgNP and a SiNP (5H) with a diameter of 300 nm and (f) two 300 nm AuNP, a 300 nm 3 SiNP (5H), and a 300 nm SiNP (20H).

The CD response of a chiral molecule can be significantly enhanced around the resonance frequency of plasmonic and dielectric nanostructures, including chiral metasurfaces[99–103], plasmonic[104–107] and dielectric nanoparticles[108,109]. These structures also allow the chiral response to be extended to the visible range, which is easy to detect and less harmful to molecules. Chiral metasufaces[99–103] allow detection of large chiral molecules of small, down to the single-molecule, concentrations. Moreover, the functionality of a wave plate can be obtained in a thin metasurface by introducing a sharp jump in the phase of the transmitted light. This can be achieved in the form of localized birefringence using anisotropic metastructures[110]. The first chiral molecule detection using a metasurface was reported in 2010 when a single-layer structure of gammadion with enhanced supra-chiral fields was used to detect various proteins[99]. The main mechanism in this work is based on measuring the spectral shift in the far-field spectrum due to interactions in the near-field between chiral molecules and the metasurface. Later, the same mechanism was extended to different chiral molecules[100–102].

Today, we can design and manufacture large-scale complex metastructures (Fig. 6c) that enhance the chiral response and facilitate the detection of chiral molecules. For instance, in ref.[113] a meta platform for chirality detection based on a plasmonic 'twisted' metamaterial has been reported. The twist angle between the particles here allows tuning of the near-field coupling between the elements that control the strength of chirality and the operating frequency, providing highly sensitive detection of the chirality of the enantiomers. The clever design of this twisted, double-layer metasurface with broken symmetry has increased the total molecular CD spectrum to several degrees, far superior to conventional CD spectroscopes, Fig. 6d,e.

Most of these works refer to structures whose chirality is fixed after fabrication. However, applications often require the ability to change the optical response, in particular chirality, and therefore require tunable and reconfigurable chiral metastructures. In a recent work[112], an all-optical atom-by-atom assembly of chiral metamolecules and in-situ measurement of their optical chirality has been reported. It is noteworthy that chiral metamolecules were disassembled into isolated meta-atoms for reorganization into their enantiomers (or other isomers), which made it possible to reconfigure chirality on demand, Fig. 6f,g. Various geometries of metamolecules have been investigated, including trimers (the simplest chiral configuration) and tetramers. The fully optical and reconfigurable assembly of chiral metamolecules presented in this study provides a simplified platform for understanding chirality at colloidal and nanoscale. All-dielectric implementation of this approach with Si NP and nanowires as building blocks in which the configuration and chiral response can be tailored on demand by dynamically manipulating the Si NP, has been reported in ref.[114].

**D. Bound states in the continuum**

Another remarkable example of the powerful opportunities that symmetry breaking provides in nanophotonics is given by the possibility of tailoring structures with very sharp resonances, of paramount importance for modern light technologies. In general, the quality (Q-) factor of any resonator is given by two contributions, the radiative Q-factor ($Q_{\text{rad}}$) due to radiation losses and the material Q-factor ($Q_{\text{mat}}$) due to Joule losses in the resonator material. Thus, the total Q-factor is defined as $Q = (Q_{\text{rad}}^{-1} + Q_{\text{mat}}^{-1})^{-1}$. In order to enhance this quantity, of interest for sharp responses, e.g., in the context of sensing and enhanced light-matter interactions, we have to reduce both loss channels. While in the RF and microwaves, the use of good metals provides low material losses due to a strong screening effect and minimal field penetration into the metal, in the optical realm, for which the value of the skin-depth is comparable to the size of metal nanostructures, the use of dielectrics with low conductivity is more preferential. Also, at sufficiently small frequencies, below 50 GHz, material losses can be significantly reduced using superconductors below critical temperature[115]. In microwaves, $Q_{mat}$ of $10^4$-$10^8$ are commonly achieved at room temperatures and up to $10^{11}$ in the superconducting regime. In optics, the design of various high-Q resonators, including microdisks[116–120], microspheres[121], Bragg reflector microcavities[122], and photonic crystals[123–125] can also provide very large Q factors ($\sim 10^3-10^6$).

However, the use of materials with low dissipative losses does not guarantee large Q factors. It is also necessary to reduce radiation losses, which is especially important in relation to open resonators, which have to be dealt with in optics. This is where the concept of embedded eigenstates (EEs), also known as bound states in the continuum (BICs), comes in. Such states have been predicted in the seminal work by von Neumann and Wigner as curious localized eigensolutions of single-particle Schrödinger equation residing within the continuum [126]. Later on, BICs have been predicted and experimentally observed in various wave settings [127]. Theoretically speaking, a BIC is a resonant state of an open system with vanishing radiative losses [128]. Due to this, BICs offer various applications, including sensing [129], lasing [130–132], and energy harvesting [133]. BICs can be classified according to the structure's symmetry and the number of involved eigenmodes (either single or multiple). The simplest example of symmetry-enabled single-mode BIC is illustrated in Fig. 7a. It is a hedgehog-like composition of elementary sources with the spherically symmetric arrangement, where the radiation is forbidden due to symmetry [134]. Since this particular case is challenging to implement, more feasible BICs in periodic 2D arrays (Fig. 7b) are of primary attention.

Systems involving multiple non-orthogonal strongly coupled modes can support Friedrich and Wintgen BICs [135]. These BICs emerge when an open cavity (Fig. 7c) supports two modes $|\psi^1\rangle$ and $|\psi^2\rangle$ that can hybridize into dressed states (Fig. 7d). Wisely tailored, one of these states can become dark, giving rise to a Friedrich-Wintgen BIC[135–137]. After terminology accepted in the literature, these modes are also called supercavity modes[138,139]. Although these states recently become a subject of intensive research in optics, especially in various photonic crystal setups[127,140–146], they are well-known in quantum physics as "dark states"[147,148]. Epsilon-Near Zero (ENZ) or other extreme material parameters allow another opportunity for BICs in multilayer structures. Here, BICs arise in structures with singular values of the permittivity[149–154] and provide versatile optical and thermal radiation properties[154,155]. This type of BIC is due to the combination of an optical resonance and a material resonance, at the ENZ point, and hence it does not directly rely on spatial symmetries. As such, it allows to avoid nanostructuring of the involved materials. Fig. 7e demonstrates the reflection spectrum for TM-polarized light from a realistic SiC (inset) structure with ENZ regime at the phonon-polariton resonance. The structure supports high Q resonances ($\sim 10^3$), limited only by material losses, and exhibits extremely narrow absorption lines for quasi-coherent and highly directional thermal radiators[154].

The non-emitting BIC state in a periodic structure like the one shown in Fig. 7b cannot be excited by a plane wave due to reciprocity. However, it is possible to introduce a broken planar inversion symmetry in order to open it and turn it into a quasi-BIC. Several typical scenarios are illustrated in Fig. 7f [146]. These scenarios can be found, for example, in refs.[156–169]. Assuming the asymmetry factor $\alpha$ to be a small perturbation, one obtains the inverse square scaling of Q:

$$Q_{rad}(\alpha) = \frac{Q_0}{\alpha^{-2}}, \qquad (2)$$

which agrees with the results of numerical simulations (Fig. 7g) and experiments [158]. Here, $Q_0$ stands for the Q-factor of the symmetric structure, $\alpha = 0$. This approach to controlled symmetry breaking for quasi-BIC excitation was studied in ref.[158], Fig. 7h, where a quasi-BIC with a record Q factor exceeding 18,000 under normal excitation was demonstrated experimentally in a Si metasurface with broken in-plane inversion symmetry.

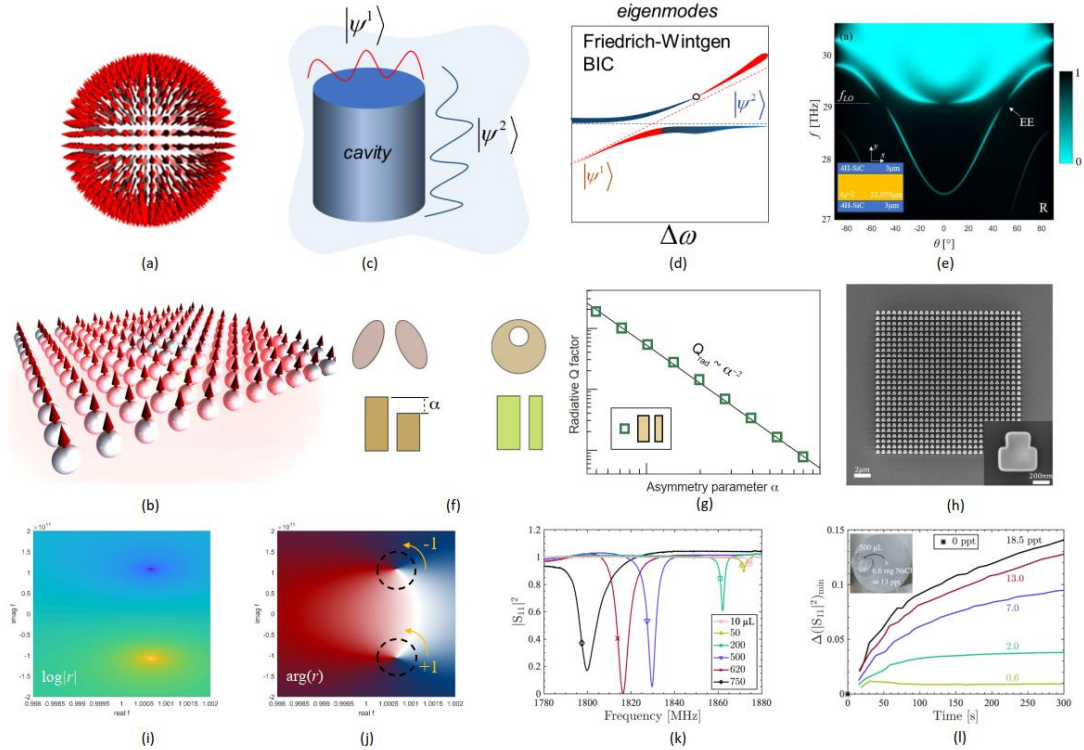

**Figure 7. Bound states in the continuum.** (a) Simplest symmetry-enabled BIC-supporting structure composed of symmetrically arranged dipole sources (red arrows). (b) BIC in periodic structures. (c,d) Illustration of an optical supercavity mode: a cavity supports two strongly coupled nonorthogonal modes $|\psi^1\rangle$ and $|\psi^2\rangle$ (c). Hybridization of the bare modes leads to the formation of two branches of hybrid (dressed) modes: one becomes dark (Friedrich-Wintgen BIC), while another becomes bright (d)[148]. (e) BIC in SiC structure with ENZ resonance. Reflection spectrum for TM-polarized light with the geometry sketched in the inset[154]. (f,g) Designs of unit cells of metasurfaces with a broken in-plane inversion symmetry of constituting meta-atoms (f). Effect of in-plane asymmetry on the radiative Q factor of quasi-BICs (g) [146]. (h) High-Q quasi-BICs in Si metasurface with broken symmetry [158]. (i,j) Amplitude and phase of the reflection around the quasi-BIC state in the complex frequency plane. (k,l) embedded eigenstate in a single resonator for advanced sensing. (k) Normalized reflection coefficient as a function of frequency and water volume. (l) Change in the reflectance minimum $\Delta(|S_{11}|^2)_{min}$ as a function of time since NaCl was added.

It has also been shown that photonic BICs in periodic structures have topological features in the form of a polarization singularities in the space of wave vectors. The robustness of these BICs was explained by their topological nature[141,144] rooted in the fact that these features obey the conservation of the topological charge[170,171]. It was also shown that the merging of BIC charges enables even more confined resonances in realistic systems[172] and unidirectional guided modes within the continuum [173]. The topological properties of the BIC have been particularly useful for polarization control, as it has been shown to be possible to convert polarization with topological protection[174–176], and circularly polarized states can arise from BICs by breaking spatial symmetries [177–179]. The generation of vortex beams at BICs [180] and efficient generation of topological vortex [181] have been recently demonstrated. A further connection was established between new topological phenomena in the form of higher-order angular states and BIC in ref.[182], indicating the far-reaching consequences that non-radiating states can have on the topological states. The topological nature of BICs in ENZ-based structures has also been discussed recently[183]. Violation of the system's Hermiticity leads to the BIC division into a pole and a conjugate zero (Fig. 7i) of opposite topological winding numbers (Fig. 7j). This reveals that perfect-absorption singularities (zeros) in lossy structures (or poles in gainy structures) are intrinsically connected to the underlying BICs.

BICs in large-area periodic arrays structures suffer from fundamental restrictions on the overall footprint and performance in the presence of inevitable disorder. The larger structure, the less feasible it for realization and use. Although localized BICs in single resonators have also been predicted [151,184,185], they exploit extreme materials suffering from losses and challenging their practical implementation. Recently, a novel approach to a BIC localized in a single subwavelength resonator based on suitable tailoring the boundaries around it has been reported[186]. The symmetry breaking of the structure has been realized by introducing a tiny water droplet atop the metallic resonator, exploiting the high permittivity of water to control the BIC state, Fig. 7k. The exciting opportunities of this boundary-induced BIC for sensing have been demonstrated by tracing NaCl dissolution in water and determining evaporation rates of distilled and saltwater with a resolution of less than 1 µL, Fig. 7l.

**E. Nonreciprocity**

Another form of symmetry breaking that has become of great interest in nanophotonics does not involve spatial symmetries, but instead time-reversal symmetry breaking. Reciprocity implies that in a system obeying time-reversal symmetry, i.e., without the presence of an external directional bias or nonlinearities, the wave transmission between any two points is independent of the propagation direction[187–195]. In particular, for any two sources $\mathbf{J}_A$ and $\mathbf{J}_B$, (Fig. 8a), the following identity applies:

$$\int_V \mathbf{J}_A \mathbf{E}_B dV = \int_V \mathbf{J}_B \mathbf{E}_A dV . \tag{3}$$

In the case of point sources, this formula yields $\mathbf{J}_A \mathbf{E}_B = \mathbf{J}_B \mathbf{E}_A$. Here, $\mathbf{E}_B$ ($\mathbf{E}_A$) is the field induced by $\mathbf{J}_B$ ($\mathbf{J}_A$) in point A (B). Nonreciprocity in bulky media requires that the electric permittivity ($\hat{\varepsilon}$) tensor is asymmetric, i.e., $\hat{\varepsilon} \neq \hat{\varepsilon}^T$ [195,196]:

$$\hat{\varepsilon} = \begin{bmatrix} \varepsilon_{xx} & i\varepsilon_\alpha & 0 \\ -i\varepsilon_\alpha & \varepsilon_{yy} & 0 \\ 0 & 0 & \varepsilon_{zz} \end{bmatrix}, \tag{4}$$

The quantity $\varepsilon_\alpha$ is the gyrotropic parameter responsible for nonreciprocal phenomena, such as Faraday rotation[197]. In many practical applications, it is advantageous to break reciprocity and time-reversal symmetry, including in the case of isolators preventing wave backscattering to the source, full-duplex systems for wireless communications that allow to transmit and receive through the same frequency channel at the same time, noise isolation in superconductor quantum computers,[198] to name a few.

Today's nonreciprocal components are almost exclusively realized based on the magneto-optical phenomena, associated with a directional bias imparted by a dc magnet. However, this approach is incompatible with planar technologies, expensive and typically associated with losses. Also, the difficulty of screening magnetic fields is harmful to superconductivity phenomena that also restricts the use of such systems in transmission-line quantum circuits[199,200]. To get around these shortcomings, one can exploit thinner or 2D materials with a magnetic response where the loss is reduced by the shorter optical path. Fig. 8b exemplifies this idea representing a dielectric (a-Si) metasurface covered by a thin layer of Ni magnetized along with the wave propagation (Faraday configuration) or along the slab perpendicular to the light wave vector (Voigt configuration). The structure encompasses two linear polarizers twisted at 45° to turn the nonreciprocal propagation phase into isolation. The metasurface can be placed on a material subject to magnetic displacement[201]. The recent discovery of magnetic atomically thin 2D materials including $CrI_3$[202], $Fe_3GeTe_2$[203], $CrGeTe_3$, $Cr_2Ge_2Te_6$[204], and $FePS_3$[205] has enabled a new class of magnetic isolators. Finally, topological insulators are also of great interest for nonreciprocity with in-build magnetic bias[206]. Fig. 8c illustrates an on-chip microwave circulator based on the quantum Hall effect nonreciprocal response. The inset shows a photograph of a circulator with a two-dimensional electron gas (2DEG) disk 330 μm in diameter (GaAs-AlGaAs interface) with a 20 μm gap to the metal[207]. The paper reports a dynamically tuned nonreciprocity of 25 dB in a 50 MHz bandwidth. A theoretical analysis of this system can be found in ref.[208].

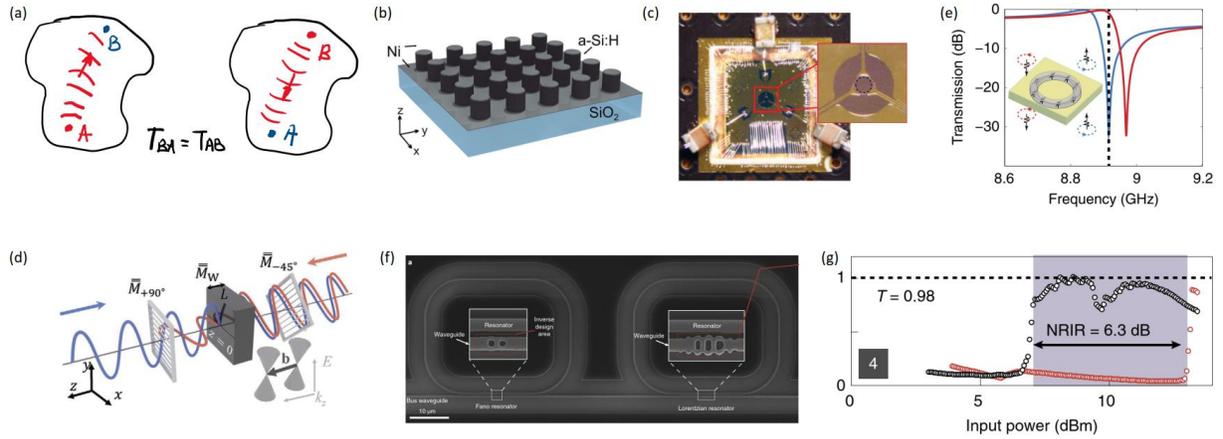

**Figure 8. Advanced nonreciprocal materials.** (a) Arbitrary structure and two current sources A and B. The reciprocity principle dictates $T(A \rightarrow B) = T(A \leftarrow B)$. (b) Schematic illustration of the magnetophotonic metasurface composed of Si nanodisks supporting magnetic Mie-type resonances. Disks are covered by a 5-nm-thick Ni film[209]. (c) Quantum Hall circulator device. Inset: Photo of the circulator showing a 330-μm diameter 2DEG disc with a 20-μm gap to the metal defining the three signal ports[207]. (e) Magnetic-free nonreciprocity with synthetic angular-momentum biasing[192]. (d) Geometry of the Faraday isolator based on Weyl semimetal slab of thickness L and Weyl nodes separation in the momentum space b. The inset shows the bulk energy E dispersion of the semimetal. The silver grids depict linear polarizers[210]. (f,g) Non-reciprocal transmission in broad operating power range using cascaded nonlinear resonators[211]. (a) SEM image of cascaded Fano-Lorentzian resonators implemented on a silicon-on-insulator platform. (g) Measured transmission versus input power of cascaded resonators in the forward (red) and backward (black) directions.

Weyl semimetals (WS), like TaAs, $WTe_2$, $TaIrTe_4$, NbP, $Co_2$-based Heusler compounds, represent a new topologically nontrivial phase of matter where the lack of time-reversal symmetry leads to massless bulk fermions and topologically protected surface states of the Fermi arc[212–215]. The band structure of Weyl semimetals contains an even number of nondegenerate Weyl nodes, the splitting of which in momentum space ensures their topological stability. The optical properties of Weyl semimetals were theoretically investigated in a number of works, for example, in refs.[216,217]. In these works, it has been demonstrated that optical properties of WS are described well by tensor Eq.(4) with the off-diagonal component[216,217]

$$\varepsilon_\alpha = \frac{be^2}{2\pi^2 \hbar \omega}.  \quad (5)$$

In ref.[210] a Faraday isolator based on a Weyl semimetal slab in the mid-IR frequency range (λ = 3.5 μm) with isolation of 41.3 dB has been proposed, Fig. 8d. It was shown that the isolation characteristics can be further improved in the geometry of a photonic crystal. There is a number of works where these topologically protected nonreciprocal effects of WS are used for unusual emission effects[218,219] and tunable superconducting quantum circuits[220].

All of these approaches are based on the use of a magnetic field, either in the form of external bias or built-in. However, there is a need for non-magnetic insulators and circulators using other approaches to electromagnetic field nonreciprocity, including space-time modulation[221–224], synthetic magnetic field[224–226], nonlinearity[227–230], interband photonic transitions[231,232], optomechanics[233–237], optoacoustics[238,239], valley polarization in transition metal dichalcogenides[240], and PT-symmetry breaking[241–243].

Magnet-free nonreciprocity based on angular momentum biasing of resonant ring metasurfaces was recently proposed[192], Fig. 8e. Modulation removes degeneracy between opposite resonance states of a metaatom. Combined with a high Q resonance, this approach provides a tremendous nonreciprocity effect. Subsequently, several applications have been proposed based on this approach, including Faraday rotation, ultrathin RF isolator, and optical isolator, all of which are implemented without the need for bulky magnetic bias elements[193]. Another approach to nonreciprocity that has been attracting much attention is based on spatially asymmetric structures and nonlinearity[244,245], ideally suited for some applications due to the lack of an external bias and its totally passive implementation. Here, reciprocity is broken due to the presence of a tailored geometrical asymmetry combined with a strongly nonlinear response. A typical implementation of this approach consists of two resonators that either have slightly different resonant frequencies or are connected differently to the circuit or free space (different $Q_{rad}$), providing an asymmetry. At least one of the resonators in this approach should be nonlinear, i.e., its resonant frequency depends on the local field intensity. These two assumptions make the resulting structure nonreciprocal[244,245]. Theoretical research and design can be performed within the framework of temporal coupled-mode theory[246]. This analysis shows that, however, this design is subject to several limitations, including dynamic reciprocity, which requires the use of these systems under pulsed operation, and a transmission-asymmetry trade-off, which requires that the maximal forward transmission ($T_{fw}$) to be very low in the regime of strong nonreciprocity and visa versa[247–249], expressed mathematically as

$$T_{fw} \leq \frac{4 \cdot \eta}{(\eta+1)^2}. \quad (6)$$

where $\eta$ is the non-reciprocal intensity range. In electromagnetically symmetric structures, $\eta = 1$ and $T_{fw}$ can reach unity but at the price of vanishing nonreciprocity. This tradeoff stems from time-reversal symmetry and it applies to devices characterized by a single nonlinear resonant system.

Nevertheless, this trade-off can be overcome by using more nonlinear resonators displaced in space. Ref.[249] reports that the combination of nonlinear Fano and Lorentzian resonators along with a suitable phase delay can provide unitary transmission with high isolation over a wide range of intensities. This idea has recently been used for nonreciprocal metasurface bilayers consisting of one nonlinear dielectric Lorentz metasurface and one nonlinear dielectric Fano metasurface.[250,251]. An interesting implementation of this nonreciprocal Lorenz-Fano response was recently proposed for chip-based LiDAR technology[211], Fig. 8f.

The system architecture has been optimized through so-called reverse engineering, which is currently used to design complex electrodynamic systems[252]. The nonlinear Lorenz-Fano system can be used to achieve fully passive nonreciprocal bias-free routing in standard silicon photonic platforms and enables performance that is not bounded by Eq. (6). As a result, large forward transmission has been reported along with a wide range of operating powers, making this design appealing for LiDAR applications, Fig. 8g. While these systems are very appealing for their passive and bias-free nature, they do not enable complete isolation under continuous-wave operations, since the superposition principle does not hold in nonlinear systems. For example, in the presence of a weak backward-propagating signal that interacts with the device together with a stronger forward wave, isolation cannot be guaranteed[253]. Rather than isolators, these nonreciprocal devices operate as nonreciprocal switches and asymmetric limiters, ideally suited for pulsed radar and LiDAR operation.

**F. PT-symmetry**

Finally, as the last class of symmetry breaking we cover in this Perspective, we discuss the physics of non-Hermitian nanophotonic systems obeying parity-time (PT)- symmetry and the optical phenomena unveiled by broken PT-symmetry. As it has been shown in the seminal work by Bender and colleagues[254,255], structures obeying PT-symmetric Hamiltonians can support a real eigenvalue spectrum $\{\omega_k\}$ in their PT-symmetric phase when the corresponding eigensolutions $\{\psi_k\}$ satisfy PT-symmetry[254]. This paradigm appeared to be very fruitful in electromagnetics and especially optics, where the existence of the PT-symmetric phase requires a balance of gain and loss. Under this condition, although energy is not necessarily conserved during their time evolution, the optical system supports a real energy spectrum. In other words, the system, on average, conserves energy in a symmetrical phase. In contrast, it can also support complex eigenvalues in the PT-symmetry broken phase with non-PT-symmetric eigenstates. When the parameter changes, the real eigenvalues change to complex ones through a second-order phase transition associated with spontaneous PT-symmetry breaking[256]. This phase transition is one of the most exciting effects of PT-symmetry as it occurs in the vicinity of exceptional points (EPs). At these transitions, the eigenvalue spectrum of these systems from real-valued becomes complex [128,254,255,257–260]. EPs are singular points in the parameter space of a non-Hermitian system at which the eigenvalues and the corresponding eigenvectors merge simultaneously. Note that these states can exist not only in 0D topology (points)[261,262] but also 1D (EP-lines and rings)[263,264] and 3D (EP-surfaces)[265]

Perhaps the simplest example of a PT-symmetric structure is a set of two coupled resonators, which are identical in every sense, except for the presence of amplification in one element (a), while the other element (b) exhibits the same amount of loss. The equation governing these systems reads $\frac{d}{dt}\begin{pmatrix}a\\b\end{pmatrix}=-i\hat{H}\begin{pmatrix}a\\b\end{pmatrix}$, where $a$ and $b$ are mode amplitudes. The Hamiltonian yields[266–270]:

$$\hat{H}=\begin{pmatrix}\omega_{0,1}-i\gamma_1 & \kappa \\ \kappa^* & \omega_{0,2}-i\gamma_2\end{pmatrix}, \qquad (7)$$

where $\omega_{0,1}$ and $\omega_{0,2}$ are the eigenfrequencies of the localized modes, $\gamma_i$ is the gain/loss magnitude in the mode $i$, and $\kappa$ denotes the coupling strength. The solution to the eigenstate problem gives the following eigenvalues ($\omega_\pm$):

$$\omega_\pm = \omega_{0,\text{ave}} - i\gamma_{\text{ave}} \pm \sqrt{|\kappa|^2 + (\omega_{0,\text{dif}} + i\gamma_{\text{dif}})^2}, \tag{8}$$

where $\omega_{0,\text{ave}} = (\omega_{0,1} + \omega_{0,2})/2$, $\omega_{0,\text{dif}} = (\omega_{0,1} - \omega_{0,2})/2$, $\gamma_{\text{ave}} = (\gamma_1 + \gamma_2)/2$, $\gamma_{\text{dif}} = (\gamma_1 - \gamma_2)/2$. In the case when $\omega_{0,1} = \omega_{0,2} \equiv \omega_0$ and $\gamma_2 = -\gamma_1 \equiv \gamma$, the eigenvalues of the two modes become $\omega_\pm = \omega_0 \pm \sqrt{|\kappa|^2 - \gamma^2}$. The analysis of eq.(8) shows that if the coupling is weaker than a certain critical value ($\kappa < \kappa_{\text{PT}} = \gamma$), the system has two modes: one with loss and one with gain. However, if the coupling is high exceeds the critical value ($\kappa > \kappa_{\text{PT}}$), the system is in the tight coupling mode, when the coherent exchange of energy between the elements compensates for the dissipation, stabilizing the system on the real frequency axis. The critical point $\kappa = \kappa_{\text{PT}}$ corresponds to the EP. Exactly at the transition threshold, $\gamma_2 = -\gamma_1$, two modes coalesce to the single eigenstate: $(1,i)^T/\sqrt{2}$ with the eigenfrequency $\omega_0$, featuring non-Hermitian degeneracy.

The abrupt phase transition at EPs enables a number of intriguing phenomena, including enhanced sensing [271–281], exotic lasing[268,282–284], optical isolation and nonreciprocity [285–288], loss-induced transparency [289], unidirectional invisibility [243,290–292], robust wireless power transfer [267,293], and topological chirality [261,294]. PT-symmetry requires a balance between loss and gain, which is difficult to achieve in experiments, especially at a higher frequency. However, some interesting physical aspects of PT-symmetry can also be investigated in non-Hermitian loss-shifted systems and loss-only structures. In fact, the Hamiltonian $\begin{pmatrix} \beta_0 - i\gamma_1 & \kappa \\ \kappa^* & \beta_0 - i\gamma_2 \end{pmatrix}$ of a lossy two-mode system can be represented as the sum of PT-symmetric ($\hat{H}_{PT}$) and decaying ($\hat{H}_L$) subsystems:

$$\hat{H} = \hat{H}_L + \hat{H}_{PT} = \begin{pmatrix} -\gamma_+/2 & 0 \\ 0 & -\gamma_+/2 \end{pmatrix} + \begin{pmatrix} \beta_0 + i\gamma_-/2 & \kappa \\ \kappa^* & \beta_0 - i\gamma_-/2 \end{pmatrix}, \tag{9}$$

where $\gamma_+ = \gamma_1 + \gamma_2$ and $\gamma_- = \gamma_2 - \gamma_1$, $\hat{H}_{PT}$ duplicates the PT-symmetric Hamiltonian (7) for balanced gain and loss rates. This approach simplifies the technical problems of realizing active PT-symmetric systems[128,289,295–297]. The EP associated with such a passive PT-symmetry is in the lower complex half-plane and can be observed either when the parameters are changed or under complex excitations[128,298].

When material changes occur on a time scale that is of the same order of magnitude as the wave period, parametric amplification is possible and strong non-Hermitianity can occur, leading to the category of time-varying non-Hermitian systems[299–301]. The use of the Floquet theorem allows the definition of the Floquet Hamiltonian, which necessarily has complex eigenvalues with a non-zero positive imaginary part. These non-Hermitian Floquet PT-symmetric systems attract much attention now because they enable various non-Hermitian effects without material gain.

Theoretically, it is possible to achieve extremely high sensitivity using the EP, since the eigenvalues around the EP have a divergent susceptibility with a small change in the parameter $\epsilon$, if the eigenvalue difference is $\delta\beta \sim \epsilon^{1/n}$, then $d(\delta\beta)/d\epsilon = \epsilon^{1/n}/\epsilon$, where $n$ is the order of EP [276,302,303]. For instance, if a system supports a second-order EP, the evolution of the eigenvalues around the EP assumes a square-root behavior superior to any conventional sensor. This conclusion, however, is still the subject of debate.

Finally, topological properties of PT-symmetric structures and EPs attract a lot of attention recently[304,305] as they enable remarkable effects such as $\pi$ Berry phase and vorticity[306,307], topological energy transfer[261,262,294], and open Fermi arcs in the bulk dispersion of systems[308].

**CONCLUSION AND OUTLOOK**

In this work, we have presented our vision of the current state of nanophotonics and the important role that broken symmetries play in the overall functionalities enabled in these systems. This emerging research area is vast and includes many different directions, with a wide range of exciting opportunities. All the classes of broken symmetries discussed in this work are promising for the future of nanophotonics. In particular, we feel that the field of photonic moiré structures is emerging and rapidly becoming an important research area. We foresee several scientific discoveries that will lead to tunable emitters, modulators, lasers, sensors and other vital devices with underlying operations driven by quasi-periodic moiré patterns. An obvious challenge for the practical application of this technology is the need for mechanical rotation, which can potentially be solved using electro-optical materials and electrical displacement in the plane. Moreover, more candidate materials and stacking scenarios need to be examined[309–311], and more approaches to reconfigurability of these devices should be explored, using, for example, phase change materials[312–315].

In addition, a significant number of theoretical and experimental advances have been reported in PT-symmetric photonics. For example, practical implementations have already reached a mature stage, going beyond the simple demonstration of PT-symmetry and its breaking. Although most of the experimental and theoretical concepts developed so far in PT-symmetric physics deal with the classical regime, we expect the expansion of these concepts to the quantum regime. Indeed, more work is needed to clarify the fundamental limitations of enhanced EP sensors and whether they can come close to the quantum noise-limited performance. One pitfall is the noise emerging due to the fluctuation-dissipation theorem that connects dissipation and gain in a system with the spectral density of noise[316]. A quantum description should encapsulate such effects[317,318]. In ref.[319], fundamental bounds on the output signal power and signal-noise ratio have been derived by analyzing a pair of coupled resonators. The results show that gain-loss PT-symmetric systems may not be beneficial for sensing applications. Considering quantum Fisher information, one can calculate the signal-to-noise ratio for EP sensors[303,320], confirming the conclusion of ref.[319]. However, the debate on the actual advantage of operating around an EP for sensing is still underway, and there is some controversy in the literature.

Interest in PT-symmetric systems in the quantum domain is not limited only to their possible application in quantum sensing. For example, a recent study[321] suggests that PT-symmetric qubits may be more resistant to decoherence and may be better suited for processing quantum information. Furthermore, a study of information flows in PT-symmetric non-Hermitian systems has shown that the complete extraction of information from the environment is possible only in the exact PT phase. Thus, phase transition at EP denotes the boundary between reversible and irreversible information flow[322].

There has been a great deal of activity associated with time-varying Hermitian and non-Hermitian systems recently. The overwhelming majority of the mechanisms used to develop topology in systems whose properties depend on time are so-called Floquet systems, where the time dependence is periodic with a modulation frequency that is usually different from the frequency of the propagating wave[323]. It is customary to distinguish between two characteristic modes: adiabatic and non-adiabatic. In the adiabatic regime, the mismatch between the modulation and wave frequencies prevents the exchange of energy between them, and the system can be considered Hermitian. The adiabatic time modulation can lead to geometric phase effects that can be used to break the time-reversal symmetry and create various topological effects[324–326]. If the modulation is fast compared to mode propagation, it is often possible to average the modulation effect, which may still have topological properties. The implementation of topological non-Hermitian time-varying systems is still in infancy[327]. Non-adiabatic temporal changes can be used to control waves[328], time modulated systems can exhibit new forms of self-amplifying wave propagation[329–331], and nontrivial topology can be obtained by modulating non-Hermitian parameters[331,332]. Geometric phase effects obtained by encircling EPs can also be observed using temporal modulations. We foresee many new capabilities when both non-Hermitisity and time modulation are applied simultaneously.

In the context of topological photonics, many problems remain unsolved. For example, a complete picture of the interplay between lattice symmetries and topological phases is steel to be obtained[333,334]. Other examples include exploiting higher-order topological phases[335] and utilizing different types of topology in applications, for example, by scaling down towards on-a-chip platforms[336]. Next, although topological systems allow reliable point-to-point energy transfer with immunity to disorder or geometric imperfections, they are inherently sensitive to absorption loss. Consequently, their expansion to the non-Hermitian regime is highly desirable. It can enable a novel form of stable topological waves[337] or make topological edge states robust to various non-Hermitian defects.

Symmetry-protected BIC states are emerging as very promising for various quantum photonics applicatons[338]. Future research in this area will be dedicated to unveiling new functionalities of these states, especially in the context of nonlinear optics. It has been recently shown that one can utilize BICs to control polarization states of light using chiral perturbations in a metasurface encoding arbitrary elliptical polarization states employing geometric phase engineering[339]. For two polarisation states, a metasurface hosting chiral BIC resonance can be resonantly coupled to one circular polarization of light while uncoupled from the counterpart enabling a narrow peak in the circular dichroism spectrum[340]. The role of symmetries and controlled symmetry breaking is particularly important in these scenarios.

Both BICs and EPs carry topological charges in their far-field radiation, but only the latter is associated with non-trivial topological invariants. Therefore, the relationship between band topology and emission topology still needs to be investigated. Another example of radiation manipulation is the realization of unidirectional guided resonances[173]. Unidirectional radiation is important for various optoelectronic applications because it can effectively improve the energy efficiency of devices such as lattice couplers[341]. Unidirectional guided resonances emerge when a pair of half-integer topological charges coalesce in momentum space and hence have a topologically nontrivial nature. These resonances are important for many optoelectronic devices, including light detection, LIDARs, and quantum cascade lasers. Thus, we can say that a huge number of various optical effects are due to the topological properties of topological charges and that the study of this area will lead to the emergence of new nontrivial effects.

Finally, time-reversal symmetry breaking is associated with the discovery of more efficient and better integrated isolators and circulators, of great interest especially in the up and coming quantum photonics domain. As we stated above, the emerging field of quantum computing is rapidly growing and has shown immense potential[342,343]. There is evidence that quantum computers hold the promise to support a computing capacity exceeding the limits of classical computers[344,345] and have the potential to efficiently solve problems currently unfeasible, drastically advancing the frontiers of high-energy physics and science in general. The building block element of quantum computing is a quantum bit (qubit). Several promising approaches to nonreciprocity compatible with fragile quantum circuits have been proposed: parametric superconducting circulators with quantum-limited frequency conversion[198,346], synthetic rotation circulators[347], and optomechanical isolators[348]. Although all these systems are magnetless, they suffer from weak forward transmission, operation and control complexity, and high energy consumption. Hence a more efficient approach to quantum nonreciprocity is yet to be proposed, and the described approaches appear appealing in this context.

All these potential directions suggest that tremendous opportunities to discover new physics and applications lie at the boundary between different approaches connected to symmetries and symmetry breaking. We believe that further progress on the physics of surface polaritons, moiré photonics, in-plane inversion symmetry breaking, nonradiative state control, time-reversal symmetry breaking and parity-time symmetry breaking, and their combinations, as discussed in this Perspective, will inspire new concepts and technologies utilizing low-symmetry nanophotonics for various applications.

**Acknowledgment**

Our body of work on these topics has been supported by the Office of Naval Research, the Air Force Office of Scientific Research, the National Science Foundation, the Department of Defense and the Simons Foundation.